# Investigation of nonlinear effects in glassy matter using dielectric methods

P. Lunkenheimer[a], M. Michl, Th. Bauer[b], and A. Loidl

Experimental Physics V, Center for Electronic Correlations and Magnetism, University of Augsburg, 86159 Augsburg, Germany

**Abstract.** We summarize current developments in the investigation of glassy matter using nonlinear dielectric spectroscopy. This work also provides a brief introduction into the phenomenology of the linear dielectric response of glass-forming materials and discusses the main mechanisms that can give rise to nonlinear dielectric response in this material class. Here we mainly concentrate on measurements of the conventional dielectric permittivity at high fields and the higher-order susceptibilities characterizing the $3\omega$ and $5\omega$ components of the dielectric response as performed in our group. Typical results on canonical glass-forming liquids and orientationally disordered plastic crystals are discussed, also treating the special case of supercooled monohydroxy alcohols.

## 1 Introduction

Despite centuries of applications of the glass transition by mankind and its general importance for different material classes, this phenomenon still belongs to the greatest mysteries in condensed-matter physics and material science. The modern definition of glass not only applies to silicate glasses used, e.g., for windows, bottles or optical components but also to the large field of polymers, metallic glasses, new types of electrolytes or even various types of biological matter (e.g., proteins). The most common way to form a glass is by cooling a liquid sufficiently fast to avoid crystallization. However, during this glass transition, glass-forming liquids continuously attain rigidity under cooling, which qualitatively differs from conventional liquid-solid transitions occurring rather abruptly at well-defined phase-transition temperatures. Moreover, the corresponding dynamics of the structural units (molecules, ions, polymer chains,...) reveals a number of universal but only poorly understood properties, which are not only relevant from an academic but also an application point of view [1,2,3,4,5,6,7].

According to the fluctuation-dissipation theorem, the measurement of susceptibilities provides direct experimental access to this dynamics. In principle, this is achieved by applying small excitations to the investigated material and monitoring its reaction. This can be done by mechanical experiments, e.g., applying constant or alternating stresses (force per unit area) to the sample and measuring the resulting deformation. This allows for the determination of mechanical moduli or compliances, which can be temperature and frequency dependent (the latter for an applied ac stress with frequency $\nu$). The probably most commonly used experimental method employed for attaining information on the molecular dynamics at the glass transition is dielectric spectroscopy. Here, usually the linear response of a glass-forming material to an applied ac electrical field is detected and presented via spectra of the complex dielectric permittivity $\varepsilon^* = \varepsilon' - i\varepsilon''$, which is directly related to the dielectric susceptibility $\chi^* = \varepsilon^* - 1$. Here $\varepsilon'$ is the dielectric constant and $\varepsilon''$ the dielectric loss, the latter characterizing the dissipation of field energy in the sample. Such experiments are intended to check the molecular fluctuations that are present even without the applied electrical field, which thus is kept sufficiently low (often a voltage of 1 V is used) to avoid any significant influence on the molecular dynamics.

---

[a] e-mail: peter.lunkenheimer@physik.uni-augsburg.de
[b] Present address: Institute for Machine Tools and Industrial Management, Technical University of Munich, 85748 Garching, Germany

However, in recent years a completely different approach is attracting increasing interest: In various works, both experimental and theoretical, it was revealed that valuable additional information on the glass transition can be gained by applying excessively high electrical fields (using voltages up to the kV range), thus driving the investigated material into a nonlinear regime (see, e.g., [8,9,10,11,12,13,14,15,16,17,18,19,20,21,22,23]). This nonlinear dielectric response is monitored, e.g., by determining the dielectric permittivity at high fields and comparing it to $\varepsilon^*$ measured in the linear regime. Another prominent method is the detection of higher harmonics of the dielectric reponse: At low field, polarization $P$ and field $E$ are proportional to each other and, thus, applying a sinus ac field $E(t)$ results in a sinusoidal sample response (quantified by the polarization $P(t)$ or the dielectric displacement $D(t)$) with the same frequency. However, at high fields, $P \propto E$ no longer is valid and an applied sinus ac field with circular frequency $\omega$ can result in higher harmonics with frequency $3\omega$, $5\omega$ etc. (no even harmonics are expected because for symmetry reasons $P(E)$ should be equal to $-P(-E)$), which is quantified by the higher-order susceptibilities $\chi_3$, $\chi_5$ etc. (see section 3.3 for a more detailed definition of quantities).

In the past, such and related experiments were used to draw far-reaching conclusions, e.g., concerning the heterogeneous nature of glassy dynamics [9,24,25]: This heterogeneity was first experimentally verified using non-linear dielectric hole-burning experiments [26] and explains one of the hallmark features of glassy dynamics, the non-exponential nature of relaxation [27,28,29]. Moreover, measurements of higher-harmonic susceptibilities revealed that increasingly cooperative motions of molecules (or ions, atoms, etc.) can explain the mysteries of the drastic continuous slowing down of molecular motion when approaching the glass transition. In this way, even strong hints on the true nature of the glass transition were obtained, which was concluded to be due to an underlying thermodynamic critical point [12,13,18].

In the present work, after a short treatment of glassy dynamics as revealed by linear dielectric measurements, we will briefly discuss the principles of nonlinear dielectric spectroscopy and possible mechanisms of nonlinearity. In the main part, we will summarize experimental results obtained by nonlinear dielectric spectroscopy. We will especially concentrate on the results from our group in the past years [16,18,21,23,30,31], which partly are the outcome of a very fruitful collaboration with the groups of F. Ladieu, G. Biroli, and J.-P. Bouchaud.

## 2 Glassy dynamics as revealed by linear dielectric spectroscopy

Figures 1 and 2 schematically indicate some of the most important properties of glass-forming liquids as detected by linear dielectric spectroscopy [7,32,33,34]. The main molecular dynamics that governs, e.g., the viscosity of a liquid is termed $\alpha$ relaxation. In dielectric spectra of the typically investigated diploar glass formers, it shows up as a step in $\varepsilon'(\nu)$ from $\varepsilon_s$ to $\varepsilon_\infty$ and a peak in $\varepsilon''(\nu)$ (Fig. 1(a)). Their shift towards lower frequencies with decreasing temperature mirrors the slowing down of molecular motion when approaching the glass transition. The step amplitude $\Delta\varepsilon = \varepsilon_s - \varepsilon_\infty$ is the relaxation strength and proportional to the molecular dipole moment and dipole density. It should be noted that dielectric spectroscopy only is sensitive to the reorientational aspect of this motion, which, however, in the vast majority of systems is rather closely coupled to the translational motions. According to the Debye theory, the loss peaks should have a half width of 1.14 decades and slopes 1 and -1 at their low and high-frequency flanks, respectively, if plotted in a double-logarithmic representation (dashed line in Fig. 1(b)) This essentially is based on the assumption of an exponential time dependence of the reaction of a molecule to external perturbations. However, nearly all glass-forming materials exhibit much broader relaxation peaks. This non-exponentiality of the $\alpha$ relaxation belongs to a number of universal non-canonical properties of glassy matter that need to be explained by any theoretical approach of the glass transition [1,4,7]. There are several empirical functions that are used to fit the experimental data. The most common ones are the Cole-Cole (CC) [35], Cole-Davidson (CD) [36] and the Fourier transform of the Kohlrausch-Williams-Watts (KWW) function [37,38] (dotted, solid and dashed lines in Fig. 1(b), respectively). Another often used formula is the Havriliak-Negami function [39]. The broadening can be explained assuming a distribution of relaxation times: Each molecule relaxes exponentially as in the Debye theory, however, with different relaxation times for different molecules. Thus, the measured broadened relaxation peaks can be assumed to be composed of numerous Debye peaks with different peak frequencies as schematically indicated in Fig. 1(c). Several key experiments, including dielectric hole burning, a special form of nonlinear dielectric experiment [26], point to the validity of this heterogeneous scenario, which nowadays is rather widely accepted (see refs. 27,28,29 for reviews on this topic).

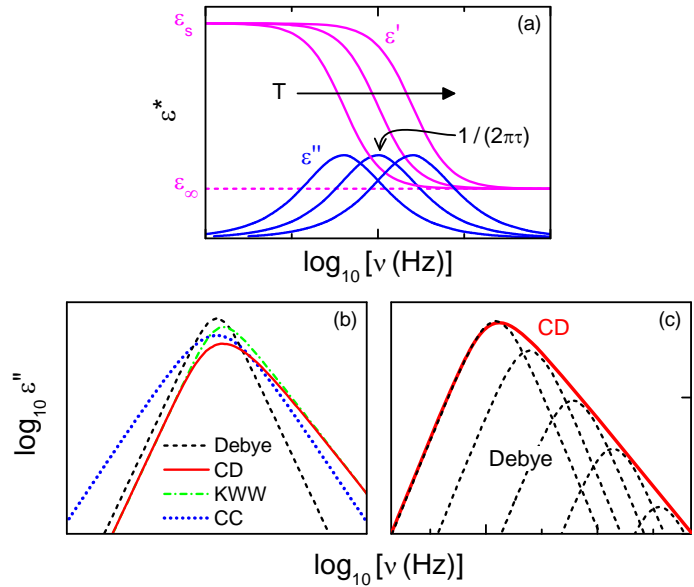

**Fig. 1.** Schematic overview of the characteristics of dielectric spectra of glass formers as revealed by linear dielectric spectroscopy. (a) Typical relaxation spectra of real and imaginary part of the permittivity, shown for three temperatures. (b) Four functions commonly used for the description of dielectric loss spectra. (c) Schematic indication of the heterogeneity-based generation of broadened loss peaks for the CD function.

Via the relation $\langle\tau\rangle \approx 1/(2\pi\nu_p)$, the frequency of the loss peaks $\nu_p$ enables a good estimate of the average relaxation time providing a measure for the molecular mobility. At the glass temperature $T_g$, originally defined as the temperature where a viscosity of $10^{12}$ Pa·s is reached, the relaxation time is of the order of 100 s. Against naive expectation, $\tau(T)$ does not follow the Arrhenius law, $\tau = \tau_0 \exp[E/(k_B T)]$ (dashed line in Fig. 2(a)), which should arise from thermally activated motion of the molecules with an activation energy $E$. Instead, for most glass formers a plot of $\log \tau$ versus $1/T$ exhibits significant curvature as indicated by the solid line in Fig. 2(a). It can be formally fitted by the empirical Vogel-Fulcher-Tammann (VFT) formula, $\tau = \tau_0 \exp[DT_{VF}/(T-T_{VF})]$ [40,41,42]. Here $D$ is the so-called strength parameter [43] and $T_{VF}$ is the Vogel-Fulcher temperature. Small values of $D$ imply strong deviations from Arrhenius behaviour. Such glass formers are also termed "fragile" (not related to fragility as a mechanical property), in contrast to so-called "strong" glass formers whose relaxation time closely follows the Arrhenius law [43]. Obviously, $\tau(T)$ would diverge at $T_{VF}$ if the VFT equation would be strictly valid (which is only rarely the case [44,45]) and can only be experimentally determined for temperatures around and above $T_g$. This divergence provides support to theories assuming that a phase transition into a state with a kind of "amorphous order" [46] may underlie the glass transition, involving diverging length scales. The non-Arrhenius behaviour is another hallmark feature of nearly all glass-forming systems [1,4,7]. It is an obvious, but controversially debated approach to trace these deviations back to a temperature-dependent apparent activation energy, strongly increasing towards low temperatures (Fig. 2(b) with inset) [1,3,4,5,47,48]. Such a behaviour may arise from an increase in cooperativity of molecular motions when the glass transition is approached [1,3,49,50]. This implies that on decreasing temperature increasingly larger regions have to move cooperatively to allow viscous flow as schematically indicated within the circles to the right of Fig. 2(b).

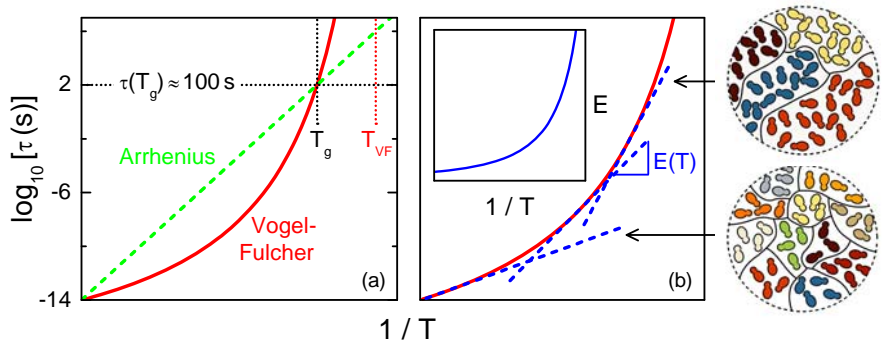

**Fig. 2.** (a) Arrhenius representation of the temperature-dependent relaxation time for Arrhenius (dashed line) and VFT behaviour (solid line). For the Arrhenius case, the activation energy is proportional to the slope in this plot. (b) A possible explanation of the non-Arrhenius behaviour of glassy matter: An increase of the size of cooperatively rearranging regions (schematically indicated by molecules of same colour at the right) leads to an increase of apparent energy barriers (inset). The latter are proportional to the slopes within the Arrhenius plot, log $\tau$ vs. $1/T$ (dashed lines).

In addition to the $\alpha$ relaxation, a number of faster dynamic processes can be detected in glass-forming matter. These processes have attracted growing interest in recent years as their understanding seems to be prerequisite to achieve a better understanding of the glass transition in general [7,33]. Figure 3 schematically shows the signatures of these processes in dielectric loss spectra for different temperatures [7,33,34]. At high frequencies, around about 1 THz, the boson peak shows up [7,34,51]. Its origin is controversially discussed and various models for its explanation have been proposed, often assuming some relation to phonon-like excitations (see, e.g., [52,53,54,55,56,57]). Above the melting point, deep in the liquid regime, it becomes superimposed by the $\alpha$-relaxation peak (Fig. 3(a)). At lower temperatures (Fig. 3(b)), the $\alpha$ peak strongly shifts to lower frequencies while the boson peak is only weakly temperature dependent. Between these two peaks, a shallow minimum shows up [7,33,34,58]. It is often assumed to signify another dynamic process, termed fast $\beta$ process or, simply, fast process and believed to arise from caged motion [59,60]. Its shallow spectral form was theoretically predicted by the mode-coupling theory of the glass transition [59]. Nonlinear dielectric measurements in the regime of the fast process or the boson peak are difficult to perform as here microwave, quasioptical THz and infrared techniques are applied, which usually do not provide sufficiently high fields to drive the material into the nonlinear regime.

Under further cooling towards $T_g$, other processes become visible in the spectra, the so-called excess wing or the Johari-Goldstein (JG) $\beta$ relaxation. In some materials, the excess wing shows up, being characterized as a second, more shallow power law at the right flank of the $\alpha$ peak [7,61], as shown in Fig. 3(c). In others, a JG $\beta$-relaxation leads to a separate peak or shoulder [62,63,64]. Both classes of glass formers are sometimes denoted as "type A" or "type B", respectively [63]. As indicated by the dash-dotted line in Fig. 3(c), it seems reasonable to assume that the excess wing is due to a JG relaxation peak that is strongly submerged under the dominating $\alpha$ peak. Indeed, there are various experimental hints in favour of this notion [65,66,67,68]. However, other scenarios regarding both spectral features as separate phenomena were also discussed [63,69]. Various explanations of the microscopic origin of the JG relaxation were suggested, e.g., motions of molecules that are located within "islands of mobility", local regions with higher molecular mobility [62]. Alternatively, the $\beta$-relaxation or excess wing were ascribed to motions of the molecules on a smaller length scale than the $\alpha$ relaxation via transitions between local energy minima arising from a fine structure of the energy landscape experienced by the molecules [70,71,72]. There are also various other approaches for the explanation of the excess wing and $\beta$ relaxation, see, e.g., [60,73,74,75].

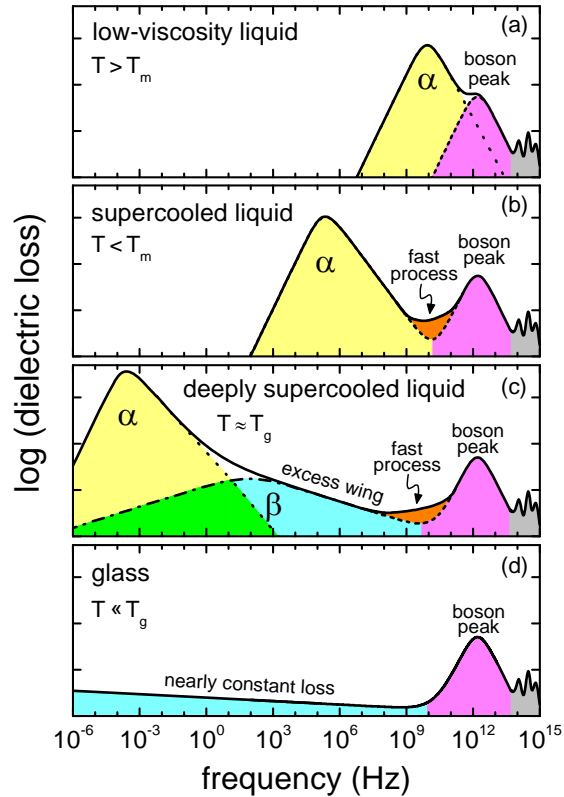

**Fig. 3.** Schematic presentation of broadband dielectric-loss spectra as observed in many glass-forming materials. In frames (a) - (d) the situations for different states of the material at different temperatures are shown: (a) Low-viscosity liquid at high temperatures, above the melting point $T_m$ [34]. (b) Supercooled-liquid regime below $T_m$ but still far above $T_g$ (c) Supercooled-liquid regime close to $T_g$ [7,33] (d) Glass, significantly below $T_g$ [34]. The contributions from the different dynamic contributions are: The $\alpha$-relaxation, the JG $\beta$-relaxation (leading to an excess wing in the present example) the fast process, the nearly constant loss, the boson peak and the infrared bands caused by intramolecular resonances at the highest frequencies. In this picture, a single secondary relaxation is assumed, but also additional ones can arise.

In the glass state, very far below $T_g$ (Fig 3(d)), the $\alpha$ and $\beta$ relaxation have slowed down so far that the corresponding loss peaks are shifted out of the frequency window. Then usually a nearly constant loss is observed over an extended frequency range. Currently, it is not clear if it is a separate phenomenon as considered, e.g., within the extended coupling model [60], or simply the high-frequency tail of the $\beta$ relaxation, which is extremely broadened at low temperatures.

## 3 Nonlinear dielectric spectroscopy

### 3.1 Different types of nonlinear dielectric measurements

Nonlinear dielectric spectroscopy can be performed in different ways: i) Application of a high ac field and measurement of the permittivity. The nonlinearity can be quantified, e.g., by the difference of high and low-field permittivity [9,16,21,24]. ii) Application of a high ac field and determination of the higher harmonics of the response, leading to higher-order permittivities [12,13,18,20,21,22,76,77]. iii) Application of a high dc bias voltage and measurement of the permittivity, usually with a smaller ac signal [78,79,80,81,82]. In addition, also hole-burning experiments can be considered as nonlinear measurements: iv) Brief application of a high ac field ("pumping"), followed by a measurement with low field [26,83]. The measurement results from such experiments are usually represented in the form of frequency and/or temperature-dependent susceptibilities as discussed in detail in section 3.3. In the present work, we concentrate on experiments according to points i) and ii) above.

It should be noted that another prominent example for nonlinear phenomena in glassy matter is found in the time-dependent variation of properties ("aging"), observed after a positive or negative temperature jump to a temperature below $T_g$ [84,85]. In such experiments, the sudden temperature jump takes the role of the field, which drives the material into non-linearity. Within the Tool-Narayanaswamy theory [86,87], this is taken into account by introducing the so-called non-linearity parameter. Several investigations of aging using dielectric spectroscopy were reported so far in literature (e.g., [88,89]). However, this type of nonlinearity is out of the scope of the present work.

### 3.2 Mechanisms leading to nonlinear dielectric response

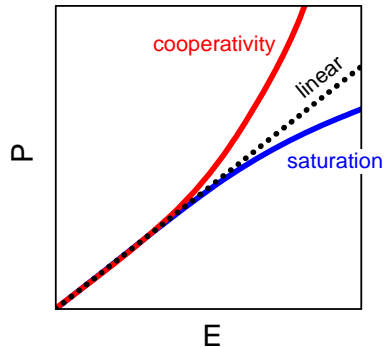

**Fig. 4.** Schematic visualization of a linear $P(E)$ behaviour (dotted line) and two possible nonlinear dependences as expected for saturation or cooperativity effects (after [96]).

A fundamental nonlinear dielectric effect was already predicted very long ago [90,91]: In the typically investigated supercooled liquids with dipolar molecules, the polarization, generated by the application of the electrical field, primarily arises from reorientational motions of the molecules. As $P^* = \varepsilon_0 \chi E^*$ with $\chi^* = \varepsilon^* - 1$, higher fields lead to stronger polarization (the stars indicate complex quantities because there can be a phase shift between $P$ and $E$). However, $P$ cannot grow infinitely and is expected to saturate when, naively spoken, all dipoles become completely oriented at very high fields (lower solid line in Fig. 4). This leads to a decrease of $\chi$, and thus of $\varepsilon$, at high fields. This saturation effect (also termed Langevin effect) is expected to be of relevance especially at low frequencies, where the field is pointing sufficiently long into one direction allowing the molecules to completely reorient. In various later works (e.g., [92,93,94]) the implications of polarization saturation on the nonlinear dielectric properties were further elaborated. However, mechanisms generating a stronger-than-linear increase of $P$ with $E$ (upper solid line in Fig. 4) were also considered early and termed "inverse saturation" or "chemical effect" [92,95,96]. In the light of more recent theoretical and experimental works [8,12,18,23], superlinear $P(E)$ behaviour may be assumed to mainly arise from cooperative motions of molecules.

For both negative and positive deviations from a linear $P(E)$ relation, when applying a sinusoidal ac field $E(t)$, a non-sinus response is expected. Via a Fourier analysis of the resulting dielectric displacement $D(t)$ or polarization $P(t)$, the higher harmonics can be obtained as schematically shown in Fig. 5, and from their amplitude and phase shift the corresponding higher-order susceptibilities can be deduced (see section 3.3).

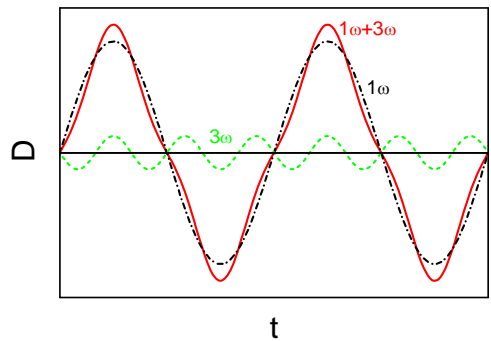

**Fig. 5.** Dissection of a non-sinusoidal signal (solid line) into $1\omega$ (dash-dotted line) and $3\omega$ components (dashed line).

As shown by Biroli, Bouchaud and co-workers [8,11], the higher-order susceptibilities are directly related to the four-point correlation function. The latter is known [97,98] to probe the cooperative length scales discussed in section 2, believed to cause the non-Arrhenius behaviour of the $\alpha$-relaxation time (cf. Fig. 2(b)). Consequently, as mentioned in section 1, various nonlinear dielectric measurements were interpreted in terms of molecular cooperativity (e.g., [12,13,18,20,21,23,31]). However, aside of the saturation and chemical effects, several other mechanisms were also considered to give rise to dielectric nonlinearity. The most prominent one is the heterogeneity of glassy dynamics, causing, e.g., the non-exponentiality of the relaxation dynamics, discussed in section 2 (cf. Fig. 1(c)) [9,10,24,25,26,30,31]. As shown by Richert and co-workers [9,24,77], within the framework of the so-called box model [26,99] a selective transfer of field energy into the heterogeneous regions can be assumed, leading to nonlinear effects in both the $1\omega$ susceptibility [9,24] and the higher harmonics [77]. Moreover, using theoretical considerations based on the Adam-Gibbs theory [50], Johari has recently demonstrated that nonlinear dielectric effects can also arise from the reduction of configurational entropy induced by the external field, leading to an increase of the relaxation time [17,100].

### 3.3 Measured quantities

Applying a time-dependent electrical field $E(t)$ to a dielectric medium generates a polarization $P(t)$. In cases where this response is nonlinear, it can be expanded as:

$$\frac{P(t)}{\varepsilon_0} = \tilde{\chi}(E)E(t) = \tilde{\chi}_1 E(t) + \tilde{\chi}_2 E^2(t) + \tilde{\chi}_3 E^3(t) + \tilde{\chi}_4 E^4(t) + \tilde{\chi}_5 E^5(t) + \ldots \qquad (1)$$

Here $\tilde{\chi}_1$ is the conventional linear part of the susceptibility $\tilde{\chi}$ and the quantities $\tilde{\chi}_i$ (with $i = 2,\ldots,\infty$) represent the higher-order susceptibilities accounting for the higher harmonics of $P$. For symmetry reasons, $P(E) = -P(-E)$ and only the odd terms of the polarization should exist. By inserting $E(t) = E_{ac}\cos(\omega t)$ for a pure ac field into Eq. (1) and using the trigonometric addition theorem, the polarization can be written as:

$$\frac{P(t)}{\varepsilon_0} = \tilde{\chi}_1 E_{ac} \cos(\omega t) + \frac{1}{4}\tilde{\chi}_3 E_{ac}^3 \cos(3\omega t) + \frac{3}{4}\tilde{\chi}_3 E_{ac}^3 \cos(\omega t) + \\ + \frac{1}{16}\tilde{\chi}_5 E_{ac}^5 \cos(5\omega t) + \frac{5}{16}\tilde{\chi}_5 E_{ac}^5 \cos(3\omega t) + \frac{10}{16}\tilde{\chi}_5 E_{ac}^5 \cos(\omega t) + \ldots \qquad (2)$$

Obviously, via this calculation $3\omega$ and $5\omega$ components automatically occur. Moreover, it should be noted that the $E^3$ component of $P(E)$ not only leads to a $3\omega$ harmonic but also to an additional $1\omega$ contribution to $P$. Correspondingly, the $E^5$ component generates additional $1\omega$ and $3\omega$ contributions. One should note, however, that Eq. (2) is oversimplified and does not take into account possible phase shifts $\delta$ between the exciting field and the polarization, arising, e.g., from the dielectric loss of the investigated material. Including $\delta$ makes the calculation more involved and leads to the following expression (with the terms now ordered by $\omega$) [13,23]:

$$\frac{P(t)}{\varepsilon_0} = \left|\chi_1^{(1)}\right| E_{ac} \cos(\omega t - \delta_1) + \frac{3}{4}\left|\chi_3^{(1)}\right| E_{ac}^3 \cos(\omega t - \delta_3^{(1)}) + \frac{10}{16}\left|\chi_5^{(1)}\right| E_{ac}^5 \cos(\omega t - \delta_5^{(1)}) \\ + \frac{1}{4}\left|\chi_3^{(3)}\right| E_{ac}^3 \cos(3\omega t - \delta_3^{(3)}) + \frac{5}{16}\left|\chi_5^{(3)}\right| E_{ac}^5 \cos(3\omega t - \delta_5^{(3)}) + \\ + \frac{1}{16}\left|\chi_5^{(5)}\right| E_{ac}^5 \cos(5\omega t - \delta_5^{(5)}) + \ldots \qquad (3)$$

Here a second index of $\chi$ is introduced: The lower one corresponds to the exponent of the electrical-field dependence while the upper one signals the $\omega$ factor ($\chi_3^{(3)}$ and $\chi_5^{(5)}$ are also often simply denoted as $\chi_3$ and $\chi_5$, respectively). In cases where a dc bias field is superimposed to the ac field, a number of new terms appear [80] which, however, are not treated in the present work. One should note that, alternatively, the higher-order susceptibilities can also be defined *including* the numerical prefactor 3/4, 10/16 etc. and care should be taken when comparing absolute values reported by different authors.

Instead of the third- and fifth-order susceptibilities, the dimensionless quantities $X_3^{(3)}$ and $X_5^{(5)}$ can be introduced [12,23]:

$$X_3^{(3)} = \frac{k_B T}{\varepsilon_0 (\Delta\chi_1)^2 a^3} \chi_3^{(3)}, \quad X_5^{(5)} = \frac{(k_B T)^2}{\varepsilon_0^2 (\Delta\chi_1)^3 a^6} \chi_5^{(5)} \qquad (4)$$

Here $\Delta\chi_1 = \Delta\varepsilon = \varepsilon_s - \varepsilon_\infty$ is the dielectric strength (cf. Fig. 1(a)).

As becomes obvious from Eq. (3), the susceptibility describing the component of $P$ proportional to $E^3$ is composed of a $1\omega$ and a $3\omega$ part ($\chi_3^{(1)}$ and $\chi_3^{(3)}$, respectively). In conventional $1\omega$ dielectric measurements at high fields, not checking for the higher harmonics, in principle $\chi_3^{(1)}$ is detected (there are also contributions from $\chi_5^{(1)}$ etc., which, however, are much smaller and usually are neglected). Usually the high- and low-field spectra of the real and imaginary part of the permittivity are recorded, which we denote here with the subscripts "hi" and "low". If defining $\Delta\varepsilon' = \varepsilon'_{hi} - \varepsilon'_{low}$ and $\Delta\varepsilon'' = \varepsilon''_{hi} - \varepsilon''_{low}$ as the differences between the high- and low-field results, $\chi_3^{(1)}$ is connected to those quantities via

$$\left|\chi_3^{(1)}\right| = \frac{4}{3}\frac{1}{E_{ac}^2}\sqrt{(\Delta\varepsilon')^2 + (\Delta\varepsilon'')^2} \quad (5)$$

and

$$\arg\left[\chi_3^{(1)}\right] = \arctan\left(\frac{\Delta\varepsilon''}{\Delta\varepsilon'}\right). \quad (6)$$

The prefactor in Eq. (5) is due to the corresponding prefactor in Eq. (3). However, instead of $\chi_3^{(1)}$, often closely related quantities are presented: For example, the vertical distance between the high- and the low-field curves of real and imaginary part of the permittivity in a logarithmic plot can be plotted, i.e. $\Delta\ln\varepsilon' = \ln\varepsilon'_{hi} - \ln\varepsilon'_{low}$ and $\Delta\ln\varepsilon'' = \ln\varepsilon''_{hi} - \ln\varepsilon''_{low}$ [9,16,21]. Alternatively, the quantity $(\varepsilon''_{hi} - \varepsilon''_{low})/\varepsilon''_{low}$ (and the corresponding one for $\varepsilon'$) can be used [81], which is comparable to $\Delta\ln\varepsilon''$ if $\varepsilon''_{hi}/\varepsilon''_{low} - 1$ is small.

### 3.4 Sample preparation and experimental details

Generally, the detection of nonlinear dielectric susceptibilities is not straightforward as the deviations from linear behaviour in the corresponding measurement signals (polarization, current etc.) are small. As the nonlinear response of a sample increases with the applied electrical field, the field amplitude should be as high as possible. Therefore, for nonlinear measurements high voltages $U$ have to be applied to the sample. Commercially available devices are usually capable of voltages extending well into the kV range. The most common sample geometry for dielectric measurements is that of a parallel-plate capacitor (Fig. 6). As here the field is given by $E = U/d$ (with $d$ the sample thickness), for reaching high fields the plate distance $d$ has to be as small as possible. For materials that are liquid at room temperature, this can be achieved, e.g., by glass-fibre spacers as schematically indicated in Fig. 6(a). In our experience, plate distances below about 30 μm are difficult to achieve in this way as electrical breakthrough probably is favoured by the sharp edges at the end of the fibres. As an alternative, shown in Fig. 6(b), silica microspheres, which are available with 1.5, 3, or 5 μm average diameter (Corpuscular Inc.), can be mixed to the sample liquid (we used mass ratios of 0.05 - 0.1 % in our measurements). Finally, liquids with sufficiently high room-temperature viscosity can also be measured without any spacer material with the upper capacitor plate "swimming" on a thin sample film (Fig. 6(c)). By carefully applying pressure to the upper plate, thicknesses of 1-20 μm can be reached in this way. The probability for electrical breakthrough is lowest for this method but sample contraction arising from the attracting forces between the capacitor plates have to be carefully excluded (see, e.g., supplementary information of Ref. [16]). In each case, highly polished plates have to be used as polishing strongly reduces the breakthrough probability. Stainless steel has proven the most suitable plate material in our experiments; the diameters of the lower and upper plates were 8 and 6 mm, respectively. The fields reached in our group using these sample preparation techniques are of the order of several 100 kV/cm [16,18,21]; the maximum field was 780 kV/cm [23]. Other setups are also possible, using different spacer materials or high-precision mechanical adjustments of the two plates to minimize their distance [9,101]. For materials that are solid at room temperature, e.g., certain plastic crystals [21,102], thin platelets have to be pressed or machined and covered by metal at both sides, e.g., by sputtering or by applying a conducting paint or paste.

Turnkey experimental setups for high-voltage dielectric measurements up to several kV are nowadays readily available. In addition, also home-made devices and setups combining various commercially available components were developed [9,12,77,103]. In our group, we are using an "Alpha-A" frequency-response analyser combined with high-voltage boosters "HVB300" with peak voltages up to 150 V or "HVB4000", all from Novocontrol Technologies. The latter is combined with a Trek 623b amplifier to reach peak voltages up to 2000 V. For the HVB4000 setup, the highest frequency is restricted to 10 kHz. Thus, whenever possible, the HVB300 booster is used ($\nu$ < 1 MHz) and sufficiently high fields are reached by minimizing the sample thickness. In addition, an "aixACCT TF2000" ferroelectric analyser with a Trek 609C-6 amplifier, enabling the application of voltages up to 1.1 kV at frequencies up to 1 kHz, is employed. Measurements with this analyser yield time-dependent polarization signals in response to the sinusoidal exciting field. A Fourier analysis of these signals and comparison with the time-dependent electrical field allows for the calculation of the dielectric permittivity and higher-order susceptibilities.

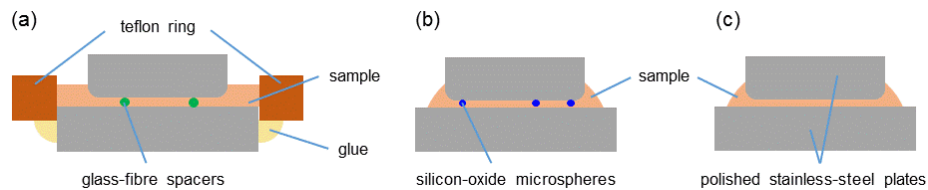

**Fig. 6.** Three capacitor setups as used by our group for nonlinear dielectric measurements.

Real-world materials have non-zero dielectric loss and conductivity and applying high voltages to lossy materials can lead to trivial heating effects, which have to be avoided in order to detect the "true" nonlinear response of a material. For this reason, the application time of the high fields has to be restricted by applying only a limited number of field cycles. Moreover, when using successive high- and low-field measurements, they should be separated by suitable waiting times. For details on used cycle successions, see, e.g., [16,24,30]. For example, in the seminal investigations reported in refs. [9,24] only five high-voltage cycles were applied, followed by a cooling period of 20 cycles. However, one should be aware that, when using a small number of cycles only, the measurement results may no longer reflect the equilibrium dielectric response [104] and a compromise has to be made aiming to minimize both heating and non-equilibrium effects.

## 4 Experimental results and discussion

**4.1 Pioneering experimental works**

The nonlinear saturation effect was first successfully detected and explained by Herweg in 1920 [90], reporting a reduction of the dielectric constant of diethyl ether with increasing field when applying voltages up to 10 kV. Debye provided further theoretical treatments of this phenomenon [91]. The first "inverse saturation effect", signifying a superlinear $P(E)$ dependence (upper solid line in Fig. 4), was found for nitrobenzene in 1936 [95]. Piekara explained this finding by the existence of coupled transient pairs of molecules [92]. The high field, favouring a parallel dipole alignment, was assumed to change the relative orientation of the molecules, leading to a small increase of the dipolar moment of the pair and, thus, a superlinear increase of $P$ with $E$. Remarkably, this explanation invokes molecular correlation effects to explain the nonlinear properties (however, for pairs of molecules only), in some respects similar to current approaches [8,12]. In this context, the so-called "chemical effect" that was reported in various, mainly chemically oriented works is also of interest (see Ref. [96] and references therein). There the electrical field is assumed to shift a chemical equilibrium towards more polar products, generating a superlinear $P(E)$ dependence. Interestingly, this concept was also applied to the self-association of molecules, which can lead to dimers or higher multimers [96]. Again this reminds of molecular cooperativity. Among the numerous further reports of nonlinear dielectric effects, finally we want to mention one of the very first investigations of the third-order susceptibility in a glass former reported by Furukawa [105]. These results on a copolymer system were described by a phenomenological model suggested by Nakada [106].

**4.2 Nonlinear 1$\omega$ measurements**

4.2.1 Canonical glass formers

In the seminal paper by Weinstein and Richert [8], high- and low-field dielectric measurements of the dielectric loss were reported for glass-forming glycerol. These results are reproduced in Fig. 7. They found a strong increase of the dielectric loss at the high-frequency flank of the $\alpha$ peak while no significant field dependence was found at lower frequencies. The authors pointed out that their results are well consistent with the heterogeneity scenario, i.e. the box model [26,99] mentioned above, assuming dynamical heterogeneities with closely correlated dielectric and thermal relaxation times: Within this scenario, the field-induced increase of $\varepsilon''$ arises from a selective transfer of field energy into the heterogeneous regions accelerating their dynamics [9,24]. The $\alpha$ relaxation in most glass formers, including glycerol [7], can be described by the CD function. The corresponding relaxation-time distribution function is strongly asymmetric [107] and there are no heterogeneous regions with relaxation rates slower than the loss-peak frequency $\nu_\alpha$ (cf. Fig. 1(c)). Therefore, only weak absorption of field energy should occur for $\nu < \nu_\alpha$ and there the loss should remain nearly unchanged for high fields as it is indeed the case in Fig. 7. Corresponding behaviour was also found for various other glass formers [16,24,25,108,109].

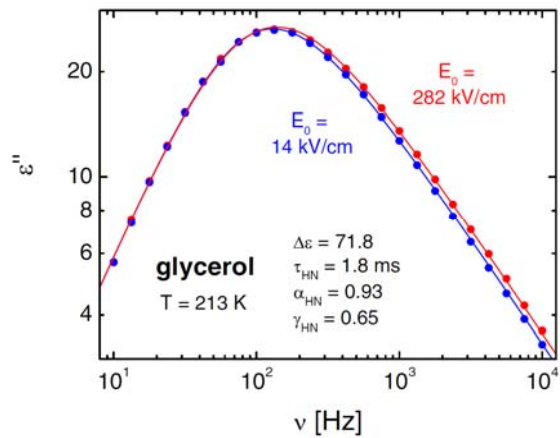

**Fig. 7.** Dielectric loss spectra of glycerol at 213 K for low and high ac field (lower and higher data points, respectively). The low-field curve was fitted by the Havriliak-Negami function [39], with the fit parameters indicated in the figure. Reprinted with permission from [9]. Copyright (2006) by the American Physical Society.

Later on, our group was able to perform corresponding experiments in an extended frequency range and at higher fields [16]. Aside of a precise determination of the nonlinearity of the $\alpha$ relaxation, this also enabled to obtain information on the nonlinear behaviour of the excess wing, a so-far mysterious spectral feature of various glass formers (cf. Fig. 3(c) and discussion in section 2). Figure 8 shows the corresponding results for the dielectric constant and loss of glycerol and propylene carbonate [16]. Surprisingly, in the spectral region of the excess wing (indicated by the dashed lines in Figs. 8(b) and (d)) we did not detect significant nonlinear effects. This becomes even clearer in Fig. 9, where the difference of the logarithmic high- and low-field loss curves is shown. For the spectra at low temperatures, where the excess wing is well within the available frequency window, $\Delta \ln \varepsilon''$ decreases towards zero for high frequencies, i.e. nonlinearity is much smaller for the excess wing than for the right flank of the $\alpha$ peak. The implications of these findings were discussed [16] within recent models on nonlinearity [8]. They seem to support long-standing assumptions of the absence of cooperativity in the molecular motions leading to secondary relaxation processes as the excess wing [110,111] (however, there are also different views [112]).

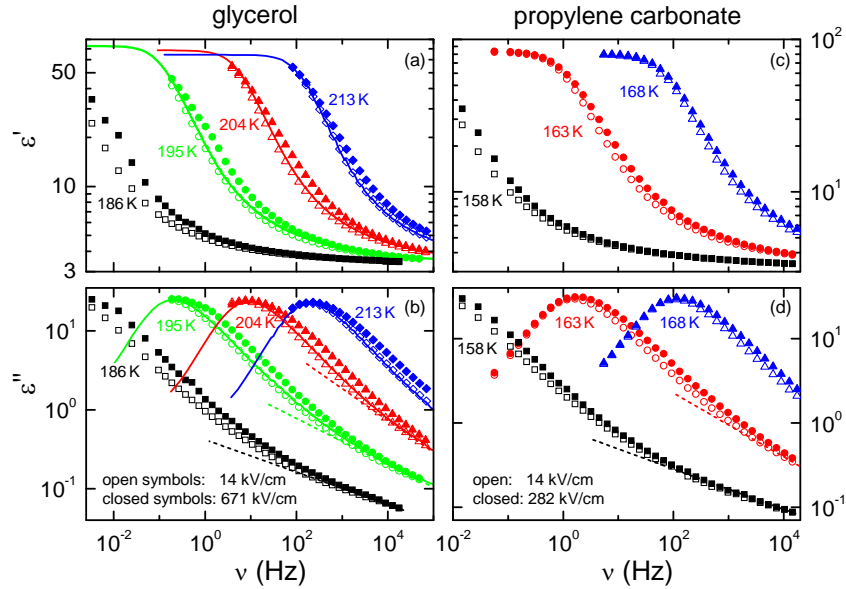

**Fig. 8.** Dielectric constant (a,c) and loss (b,d) of glycerol (a,b) and propylene carbonate (c,d) [16]. Open and closed symbols denote results for low and high electrical fields, respectively, as indicated in the figure. The solid lines in frames (a) and (b) show the low-field results from Ref. [58], measured with 0.2 kV/cm, which reasonably agree with the data for 14 kV/cm. The dashed lines in frames (b) and (d) indicate the excess wing.

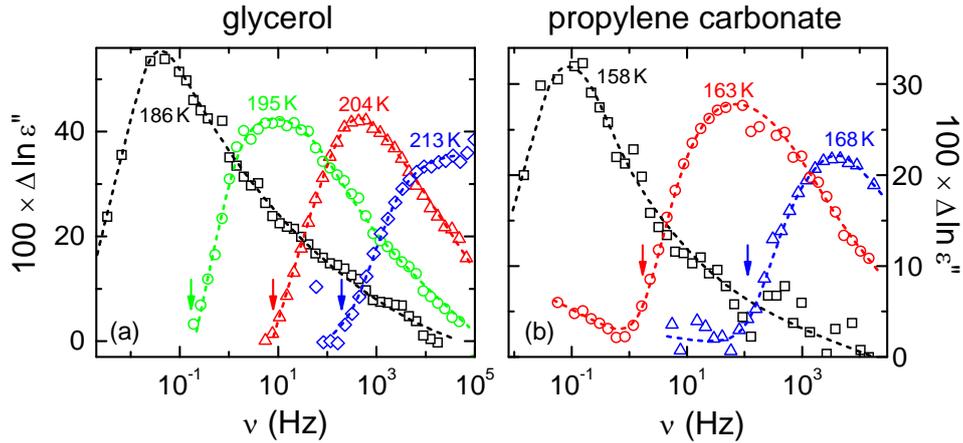

**Fig. 9.** Difference of the logarithm of the loss spectra of glycerol (a) and propylene carbonate (b), measured with high fields of 671 kV/cm (glycerol) or 282 kV/cm (propylene carbonate) and with a low field of 14 kV/cm. The arrows indicate the $\alpha$-peak positions. The lines are guides to the eyes.

In the measurements of Ref. [16], following the procedure described in [9,24] we took special care to safely exclude trivial heating of the sample by using a succession of high- and low-field cycles ensuring only a limited number of applied high-field cycles (see section 3 and supplementary information of Ref. [16]). Later on, Samanta and Richert performed time-resolved measurements of the loss at high fields [104]. Interestingly, they found that, when applying the high field for up to several 10000 cycles, an equilibrium state is reached and, in this way, nonlinearity can also be detected in the excess-wing region. In any case, the nonlinearity in this region reported in Ref. [104] is also clearly smaller than for the main relaxation and the overall nonlinear behaviour of the excess wing significantly differs from that of the $\alpha$ relaxation (see also [113] for a detailed discussion of these issues).

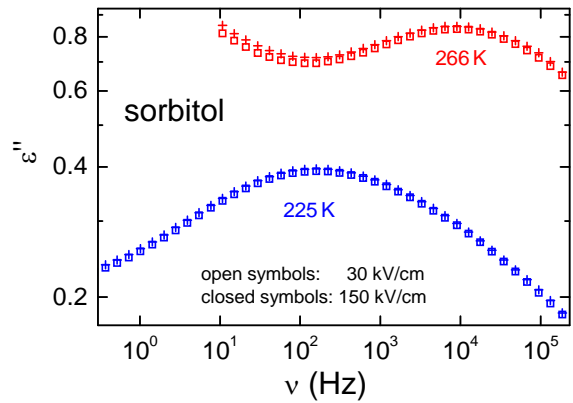

**Fig. 10.** Dielectric loss of sorbitol at two temperatures in the reagion of the $\beta$ relaxation [114]. Squares and crosses denote results for low and high electrical fields, respectively, as indicated in the figure.

While glycerol and propylene carbonate are type A glass formers, exhibiting an excess wing, Samanta and Richert [115] have also investigated the nonlinear dielectric response of the type-B glass former sorbitol, which is known to show a well-pronounced $\beta$ relaxation [64]. In the region of the $\beta$-relaxation peak, they find a field-induced increase of the dielectric loss. Similar results from Ref. [114] are shown in Fig. 10. It seems that the whole $\beta$-relaxation peak is shifted upwards by the high field, which corresponds to an increase of the relaxation strength. For 266 K, a minimum is seen at low frequencies due to the onset of the $\alpha$-relaxation peak. At frequencies below this minimum, an even stronger increase of the loss is detected, corresponding to the nonlinearity effects of the $\alpha$ peak known form other glass formers [9,16] (Fig. 8). Our sorbitol measurements were done with a small number of cycles, just as our investigations of the excess wing [16], to exclude trivial heating effects [114]. In contrast to the excess wing in type A systems, even with small cycle numbers significant nonlinear effects are detected for the $\beta$ relaxation of sorbitol. However, in Ref. 115, applying many cycles, very similar nonlinear behaviour of excess wing and $\beta$ relaxation were reported and it was concluded that both phenomena have the same origin as indicated in Fig. 3(c) [65,66,67,68]. Obviously, further type B systems have to be investigated to clarify possible universalities in their nonlinear behaviour. In this context, it should be noted that the nature of the $\beta$ relaxation in sorbitol is not so clear. In [116,117] it was identified as a "genuine" JG $\beta$-process, while the results in Refs. [64] cast some doubts on this finding. Finally, we want to mention that the small or even absent nonlinear effects in the excess-wing region and the contrasting clear nonlinearity found for the $\beta$ relaxation in sorbitol were consistently explained within the framework of the coupling model [60], where the excess wing is identified with the so-called "nearly constant loss" caused by caged molecular motions [113].

4.2.2 Plastic crystals

As schematically indicated in the inset of Fig. 11, in plastic crystals the molecules are arranged in a well-defined crystalline lattice but their orientational degrees of freedom are still disordered. The latter can exhibit glassy freezing, which in linear dielectric spectroscopy leads to very similar phenomenology as in canonical glass formers [102]. Only recently, nonlinear dielectric spectroscopy was also reported for plastic crystals [21,31,118,119]. Figure 11 shows the nonlinear $1\omega$ response of two plastic-crystalline materials, cyclo-octanol [21] and a mixture of 60% succinonitrile and 40% glutaronitrile (60SN-40GN) [31]. Again, the logarithmic difference of the high- and low-field results is plotted. For cyclo-octanol, $\Delta \ln \varepsilon'$ exhibits a succession of a negative and positive peak, which is accompanied by a V-shaped behaviour of $\Delta \ln \varepsilon''$, both centred around the $\alpha$-peak frequency indicated by arrows (Figs. 11(a) and (b)). As noted in ref. [21], this corresponds to a broadening of the $\alpha$-relaxation step or peak at frequencies both above *and* below the peak frequency. In contrast, in canonical glass formers only weak nonlinear $1\omega$ effects are observed at low frequencies [9,18,24].

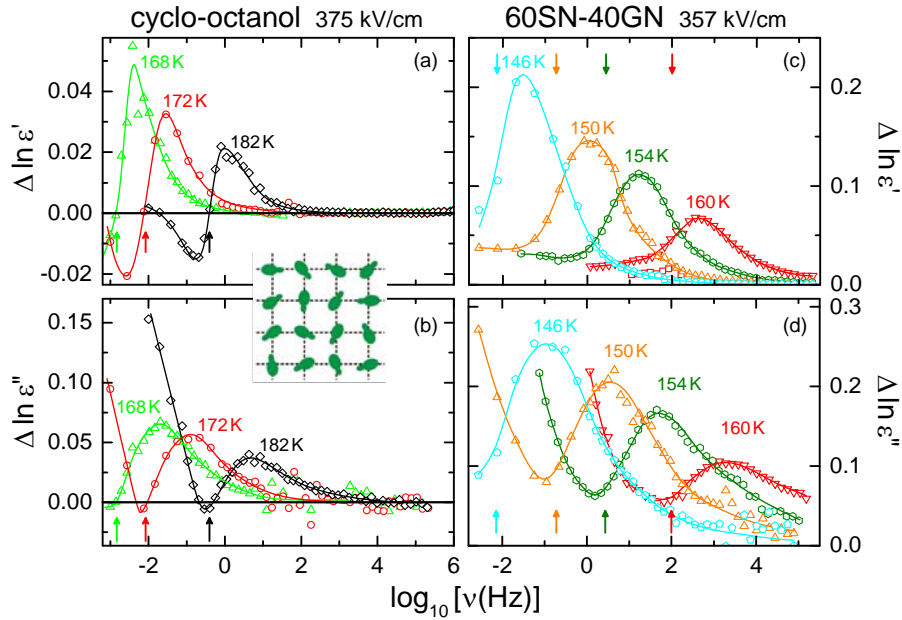

**Fig. 11.** Difference of the logarithm of the loss spectra of plastic-crystalline cyclo-octanol (a,b) and 60SN-40GN (c,d), measured with high fields of 375 kV/cm (cyclo-octanol) or 357 kV/cm (60SN-40GN) and with a low field of 14 and 13 kV/cm, respectively [21,31]. The lines are guides to the eyes. The arrows indicate the $\alpha$-peak positions. The inset schematically indicates the orientational disorder of plastic crystals.

The field-induced increase of $\varepsilon'$ and $\varepsilon''$ at frequencies $\nu > \nu_\alpha$ can be explained [21,31,118] by a similar mechanism as in structural glass formers, namely a selective transfer of field energy into the heterogeneous regions, accelerating their dynamics [9]. According to Ref. [118], the *low*-frequency nonlinear effects in plastic crystals may be ascribed to the reduction of configurational entropy induced by the external field, which should lead to an increase of the relaxation time. This notion arises from recent theoretical considerations by Johari [17,100], based on the Adam-Gibbs theory [50]. This entropy effect should be most pronounced at low frequencies because the molecular arrangements associated with the entropy reduction are too slow to lead to significant effects at high frequencies [118]. Interestingly, entropy-driven nonlinearity seems to be more pronounced in plastic crystals [21,31,118] than in structural glass formers [9,16,24,119]. This effect is obviously based on a field-induced modification of the reorientational degrees of freedom of the molecules [17]. As in plastic crystals molecular reorientations are much more important for the overall entropy than in structural glass formers, where additional translational degrees of freedom exist, one may speculate that this is the reason for the different behaviour of these two classes of glassy matter [31].

Cyclo-octanol exhibits two relaxation processes faster than the $\alpha$ relaxation, termed $\beta$ and $\gamma$ relaxation [120]. In the spectral regions dominated by these high-frequency relaxations, the field-induced variation of $\varepsilon'$ or $\varepsilon''$ is small and the logarithmic differences approach zero at high frequencies (Figs. 11(a) and (b)). This resembles the reduced nonlinearity in the excess-wing region of canonical type-A glass formers [16] (Figs. 8 and 9) but contrasts with the behaviour of the type-B glass former sorbitol (Fig. 10) [115,114].

The plastic-crystalline mixture 60SN-40GN [121] exhibits qualitatively similar nonlinear behaviour of $\varepsilon''$ (Fig. 11(d)) as cyclo-octanol (Fig. 11(b)) [31]: A V-shaped feature with the minimum close to the corresponding $\alpha$-peak frequency (arrows) is followed by a continuous decrease towards high frequencies. Similar effects as in cyclo-octanol can be invoked to explain these findings. However, at the minimum $\Delta \ln \varepsilon''$ does not approach zero and $\Delta \ln \varepsilon'$ exhibits a low-frequency plateau instead of a negative peak. As discussed in detail in Ref. [31], both findings correspond to a field-induced increase of the static dielectric constant $\varepsilon_s$. It can be explained if considering the fact that both succinonitrile and glutaronitrile molecules can assume different conformations with different dipolar moments [122,123]. Therefore, the applied high electrical fields may favour conformations with higher dipolar moment, leading to an increase of the average dipolar moments of the mixture thus explaining the detected enhancement of $\varepsilon_s$. This is another mechanism that can lead to nonlinear dielectric effects and can be regarded as a special type of "chemical effect" as discussed in Ref. [96].

4.2.3 Monohydroxy alcohols

Glass-forming monohydroxy alcohols are known to exhibit a peculiarity in their dielectric spectra, namely a Debye-type relaxation process that is slower, i.e. located at lower frequencies, than the $\alpha$ relaxation governing, e.g., viscous flow [124,125]. This process is nowadays quite commonly ascribed to the relaxation of clusters formed by several hydrogen-bonded alcohol molecules [14,125,126]. The Debye relaxation of monohydroxy alcohols represents an idealised case of a relaxation process as it lacks the broadening commonly found for other glass formers. Thus, heterogeneity, which can cause nonlinear effects, should not play a role. Moreover, cooperativity, also considered to give rise to nonlinearity, should be less important, too, because cluster-cluster interactions should be rarer than the intermolecular interactions in other glass formers [30].

Figure 12 presents the low- and high-field dielectric spectra for 1-propanol [30]. The Debye relaxation leads to a well pronounced step in $\varepsilon'$ (Fig. 12(a)) and peak in $\varepsilon''$ (Fig. 12(b)) while the $\alpha$ and $\beta$ relaxations in this alcohol only show up as weak shoulders [127]. The insets of Fig. 12 show the logarithmic differences of the high- and low-field spectra. In Ref. [30], the strong reduction of $\varepsilon'$ at low frequencies under high field, leading to negative $\Delta \ln \varepsilon'$ and correspondingly negative $\Delta \ln \varepsilon''$, was ascribed to the saturation effect (see section 3.2). The magnitude of this reduction is consistent with the cluster-like nature of the relaxing entities and enabled conclusions concerning the cluster size [30]. The increase of $\varepsilon'$ and $\varepsilon''$ at the high-frequency flank of the Debye peak was ascribed to the absorption of field energy as considered in [25] for the monohydroxy alcohol 2-ethyl-1-butanol. The behaviour in the $\alpha$-relaxation region was explained along similar lines as for canonical glass formers [9,16], invoking heterogeneity effects. Finally, for the $\beta$ relaxation no significant nonlinear contributions were detected. It should be noted that similar but partly also quite different nonlinear behaviour was reported for other monohydroxy alcohols [14,25,128]. Of special interest is the finding of nonlinearity due to transitions between less polar ring-like cluster structures and more chain-like structures, the latter being preferred under high fields due to their higher dipolar moments [128]. 1-propanol does not seem to be affected by such effects.

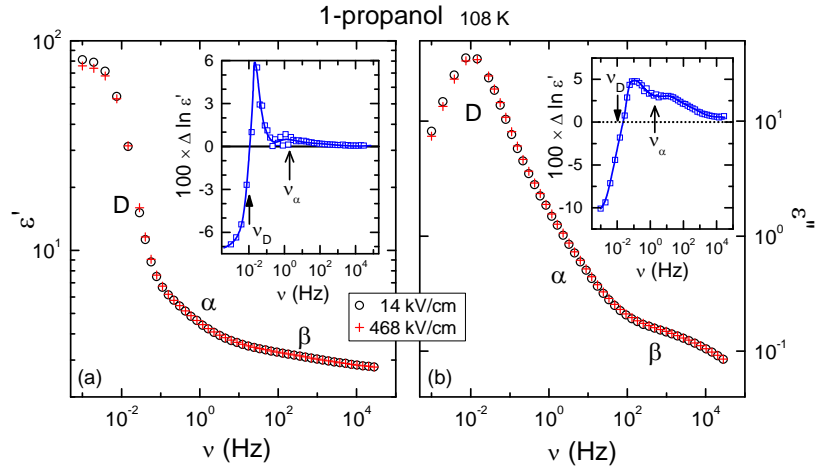

**Fig. 12.** High- and low-field spectra of $\varepsilon'$ (a) and $\varepsilon''$ (b) for glass-forming 1-propanol at 108 K. The insets show the logarithmic differences with the Debye- and $\alpha$-peak frequencies indicated by arrows. The lines are guides to the eyes.

**4.3 Higher harmonics**

4.3.1 Third harmonic dielectric response

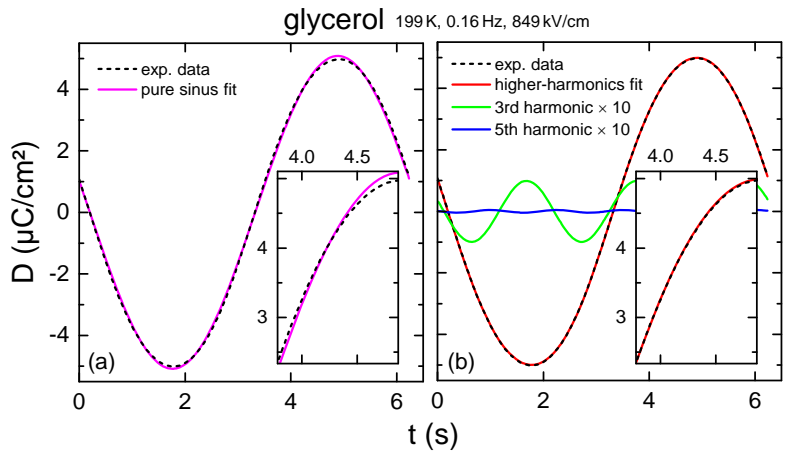

**Fig. 13.** Time-dependent dielectric displacement of glycerol at 199 K, 0.16 Hz and 849 kV/cm (dashed lines) [114]. The solid line in (a) is a fit with a single sinus function. In (b) the solid line superimposed to the experimental data is a fit with three sinus components ($1\omega$, $2\omega$ and $3\omega$). The other solid lines show the $3\omega$ and $5\omega$ contributions, multiplied by a factor of 10. The insets show a magnified view of the region preceding the maximum of $D(t)$.

To demonstrate the contributions of the higher harmonics to the dielectric response, Figure 13 shows the dielectric displacement as measured for glycerol at 199 K [114]. Frame (a) reveals that a fit of the experimental data (dashed line) with a single sinus function (solid line) leads to small but significant deviations (see also inset). Only a fit function combining 1ω, 3ω and 5ω components, i.e. including higher harmonics, leads to a perfect fit of the data (frame (b)). It should be noted that the higher-harmonic components, also shown in Fig. 13(b) (note their scaling by a factor of 10), are very small compared to the $1\omega$ contribution, which makes the detection of the higher-order susceptibilities a difficult task, implying high requirements for the precision of the used experimental setup.

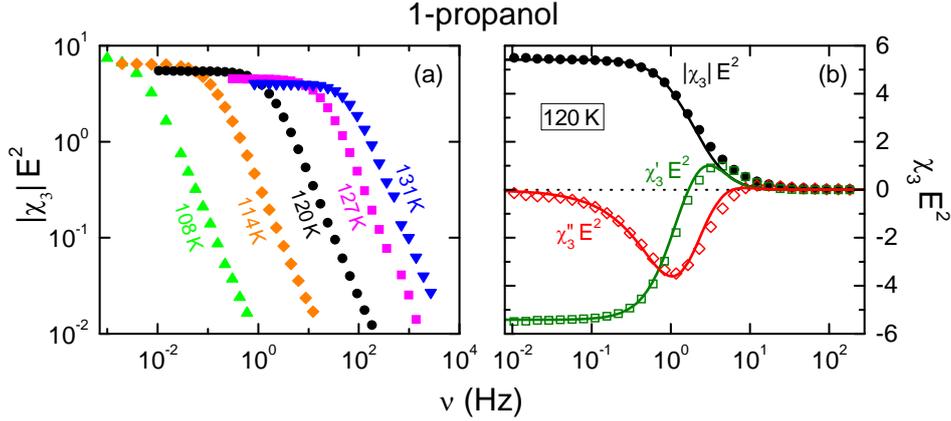

**Fig. 14.** (a) Modulus of the third-order dielectric susceptibility (times $E^2$) of 1-propanol for various temperatures as measured with a field of 468 kV/cm. (b) Modulus, real and imaginary part of the same quantity at 120 K [114]. The solid lines were calculated according to Ref. [94].

As discussed in section 3.2, the simplest nonlinear dielectric effect is caused by polarization saturation at high fields. As this leads to deviations from a linear $P(E)$ dependence (Fig. 4), higher-order susceptibilities should arise. Especially, spectra of $\chi_3$ should exhibit a plateau at low frequencies, followed by a continuous decrease with increasing frequency [23,94] The Debye relaxation in monohydroxy alcohols seems best suited to check for such behaviour as it is expected to be not affected by additional heterogeneity or cooperativity-induced nonlinear contributions (see discussion in section 4.2.3). Figure 14(a) shows the modulus of $\chi_3 E^2$ which indeed follows the expected behaviour (we plot $\chi_3 E^2$ instead of $\chi_3$ as this is a dimensionless quantity, just as the linear permittivity; cf. Eq. (3)). In Fig. 14(b), the modulus, real and imaginary parts of this quantity are shown as measured at 120 K [114]. The solid lines were calculated with the formulae provided by Dejardin and Kalmykov accounting for the saturation effect [94]. A good qualitative agreement of calculated and experimental data can be stated when using the correct values for the peak frequency, relaxation strength and molecular volume for the calculation and applying an additional factor of 2.9. This additional factor is in accord with the notion that the Debye relaxation in the monohydroxy alcohols reflects the dynamics of clusters formed by several molecules [14,125,126]. The deviations of the calculated and measured spectra, becoming most obvious at high frequencies, may arise from the absorption of field energy or the influence of the $\alpha$ relaxation.

Figure 15 shows the modulus of $\chi_3 E^2$ for three different types of glass formers, namely glycerol (canonical supercooled liquid), 2-ethyl-1-hexanol (2E1H; monohydroxy alcohol) and cyclo-octanol (plastic crystal) [18,21]. In all three cases, a hump is observed at a frequency somewhat below that of the main peak in the linear dielectric loss spectra. Such a humped spectral shape of $\chi_3$ was first observed in the pioneering work by Crauste-Thibierge *et al.* [12] for glycerol and our results in Fig. 15(a) nicely agree with those reported there. Within the model by Biroli and co-workers [8,11], such behavior is predicted to arise from molecular cooperativity, typical for glass-forming systems. It is an interesting finding that the alcohol 2E1H (Fig. 15(b)) also exhibits such a hump, in contrast to 1-propanol (Fig. 14), where the trivial saturation effects seem to be sufficient to explain the general trend of the $\chi_3$ spectra. This is consistent with the fact that monohydroxy alcohols exhibit a large diversity concerning their nonlinear dielectric behaviour [14,25,30,128]. Finally, the detection of such a hump in cyclo-octanol (Fig. 15(c)) indicates that the glassy freezing in plastic crystals is also governed by molecular cooperativity. In [21] it was proposed that lattice strains, reducing the energy barriers for reorientational motions of neighbouring molecules, are the main mechanism for generating molecular correlations in plastic crystals.

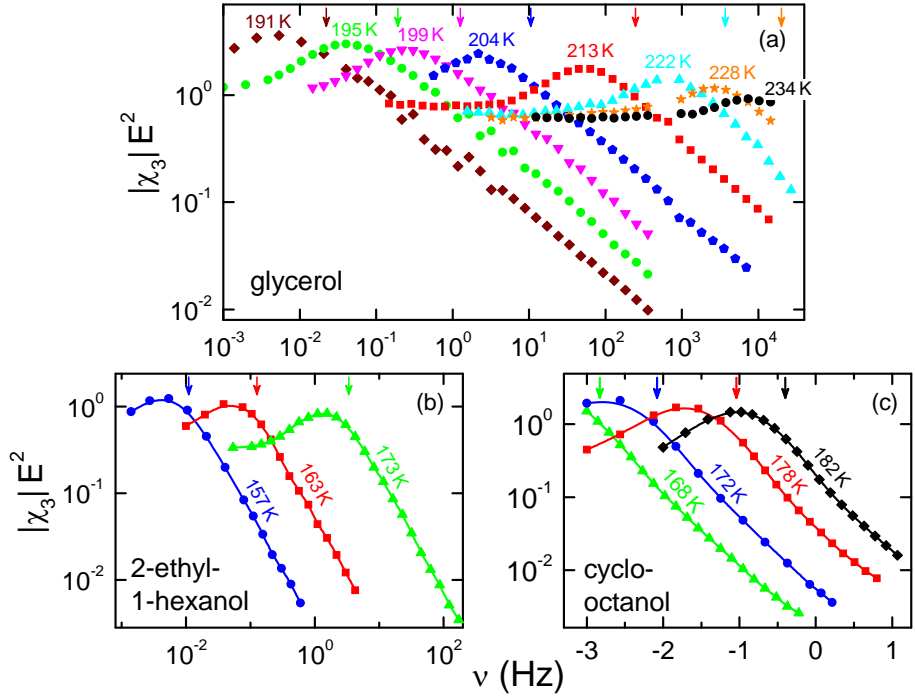

**Fig. 15.** Modulus of the third-order harmonic component of the dielectric susceptibility (times $E^2$) of glycerol [18] (a), 2E1H (b) [18] and cyclo-octanol [21] (c), measured at various temperatures. The applied fields were 565 kV/cm for glycerol, 460 kV/cm for 2E1H and 375 kV/cm for cyclo-octanol. The arrows indicate the $\alpha$-peak frequencies. The lines in (b) and (c) are guides for the eyes.

In Fig. 16, for glycerol in addition to the modulus, the real and imaginary parts of the third-order susceptibility (times $E^2$) are presented, too [114]. As shown by the lines [94], the saturation effect alone is unable to provide even a qualitative description of the experimental data. Only at the lowest frequencies, the occurrence of plateaus in the experimental curves agrees with the trivial expectation. It seems reasonable that, on very long time scales, the liquid flow (directly related to the $\alpha$ relaxation) destroys glassy correlations and the trivial saturation effect dominates [23]. Finally, it should be mentioned that in several works it was pointed out that humped $\chi_3$ spectra as shown in Fig. 15 can also arise from other mechanisms than cooperativity [15,19,76,77,129,130]. However, to us the cooperativity-related mechanism seems the most reasonable explanation, as it is also consistent with the observed variation in the non-Arrhenius behaviour of the temperature-dependent relaxation times of the investigated glass formers as discussed in the following.

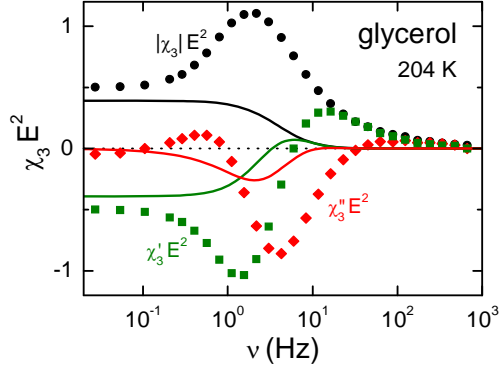

**Fig. 16.** Modulus, real and imaginary part of $\chi_3 E^2$ of glycerol at 204 K, measured for a field of 354 kV/cm [114]. The solid lines were calculated according to Ref. [94] and provide an estimate of the trivial $3\omega$ response of glycerol.

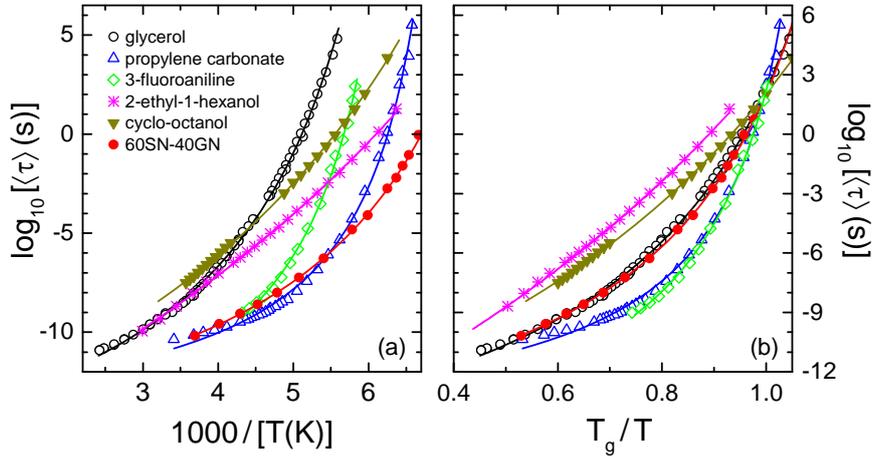

**Fig. 17.** (a) Temperature-dependent average relaxation times of the glass formers, for which humps in $\chi_3$ were detected, shown in Arrhenius representation [45,63,120,121,125]. The lines are fits with the VFT function. (b) Same data shown in an Angell plot [131]. $T_g$ was determined by the condition $\langle\tau\rangle(T_g) = 100$ s, except for the monohydroxy alcohol 2E1H, where this condition was applied to the $\alpha$ relaxation instead of the shown $\langle\tau\rangle$ of the Debye relaxation.

The temperature-dependent average relaxation times of the glass formers for which humps in $\chi_3$ were detected by us are plotted in Fig. 17(a) using an Arrhenius representation. Different degrees of deviations from straight-line behaviour, which would signify simple thermally-activated Arrhenius temperature dependence, are revealed. This becomes even better obvious in the Angell plot, showing the same data versus a temperature axis that is scaled by $T_g$ [131]. Propylene carbonate [45] and 3-fluoroaniline [63] are the most fragile systems while glycerol [45] and 60SN-40GN [121] are so-called intermediate glass formers. Cyclo-octanol [120] and 2E1H [125] behave rather strong. Here it should be noted that for the monohydroxy alcohol 2E1H, the Debye relaxation is shown, which is the relevant relaxation when discussing its nonlinear $\chi_3$ behaviour (see below). For a discussion of its $\alpha$ relaxation, see, e.g., [125]. The strong (i.e., non-fragile) $\tau(T)$ characteristics of cyclo-octanol is in accord with the findings for most other plastic crystals [102,132]. In contrast, 60SN-40GN can be regarded as an unusually fragile plastic crystal which was ascribed to the additional substitutional and conformational degrees of freedom in this system [121].

Can the markedly different temperature characteristics of these glass-forming systems be related to different behaviour of their molecular cooperativity as suggested by Fig. 2(b)? Within the model by Biroli, Bouchaud and co-workers [8,11], the non-trivial part of the third-order susceptibility $\chi_3$ (i.e., the hump observed in Fig. 15) should be related to the number of correlated molecules $N_{corr}$. Especially, the dimensionless quantity $X_3^{(3)}$, defined in Eq. (4), which is corrected for trivial temperature dependences, should be directly proportional to $N_{corr}$. In Fig. 18, we show the peak value of $X_3^{(3)}$, i.e. $N_{corr}$ in arbitrary units, versus temperature for the investigated materials (symbols; left scale). As already found for glycerol by Crauste-Thibierge *et al*. [12], $N_{corr}$ for all systems increases with decreasing temperature. This implies a growth of correlation length scales, consistent with a phase-transition related origin of the glass transition. Interestingly, marked differences in the temperature variation of $N_{corr}$ show up: propylene carbonate and 3-fluoroaniline obviously have the strongest temperature dependence, while the temperature-dependence is weakest for 2E1H and cyclo-octanol. $N_{corr}(T)$ of 60SN-40GN and glycerol behave intermediate between these two extremes. Obviously, these differences are fully correlated with the different fragilities of these six systems as noted in the preceding paragraph (Fig. 17): Propylene carbonate and 3-fluoroaniline are fragile, which implies strongly temperature-dependent, super-Arrhenius $\tau(T)$ behaviour. 2E1H and cyclo-octanol are strong, i.e. their $\tau$ has much weaker temperature dependence, and 60SN-40GN and glycerol are intermediate. These findings strongly corroborate the notion that $N_{corr}$ determined from $\chi_3$ is a meaningful quantity that governs the relaxation dynamics of glass-forming materials.

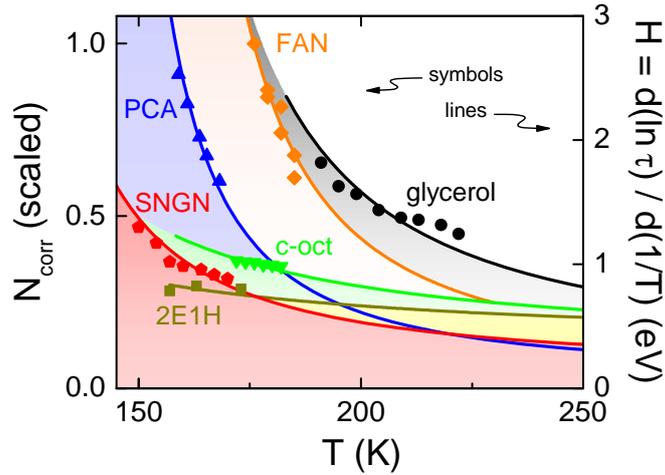

**Fig. 18.** Comparison of activation energies and $N_{corr}$ [18,21,31]. The lines indicate the effective activation energies $H$ determined from the derivatives of the temperature-dependent relaxation-times (right scale). The symbols (left scale) show $N_{corr}$, the number of correlated molecules as determined from $\chi_3$. It is shown in arbitrary units. $N_{corr}$ was multiplied by separate factors for each material (glycerol: 1.15, propylene carbonate (PCA): 0.72, 3-fluoroaniline (FAN): 1.30, 2E1H: 0.39, cyclo-octanol: 0.19, 60SN-40GN: 1.05), which leads to a good match with the derivative curves. The fact that both ordinates start from zero implies direct proportionality of both quantities.

Aside of this clear qualitative connection of $\tau$ and $N_{corr}$, is it also possible to find a quantitative relation? In the light of the energy-barrier-related explanation of the non-Arrhenius behaviour of $\tau(T)$ as indicated in Fig. 2(b), fragile behaviour implies a strong temperature variation of the energy barriers $E$, governing molecular motion, while strong behaviour corresponds to temperature-independent $E$. Such a connection of $\tau$ and $E$ was already considered in one of the most prominent early theories of the glass transition, the Adam-Gibbs theory [50]. There it is assumed that the size of cooperatively rearranging regions grows when approaching the glass temperature under cooling and that the "cooperative transition probability" is thermally activated with a temperature-dependent energy barrier that is proportional to the number of molecules within such a region. This essentially suggests that $E \sim N_{corr}$. As indicated in Fig. 2(b), temperature-dependent apparent energy barriers can be estimated from the experimental $\tau(T)$ data by determining their derivative in the Arrhenius representation, $H = d(\ln\tau) / d(1/T)$ (we used derivatives of the fit curves instead of the experimental data to avoid excessive data scatter). The results are presented by the lines included in Fig. 18 (right scale). After applying suitable scaling factors, the $N_{corr}(T)$ data (which is given in arbitrary units) as deduced from $\chi_3$ match nicely the $H(T)$ curves, the latter well reproducing the markedly different temperature variations of $N_{corr}$ of the investigated materials. The fact that both ordinates in Fig. 18 start at zero implies that, indeed, the activation energy governing $\tau(T)$ is approximately proportional to $N_{corr}$.

The factors used to make $N_{corr}$ match the corresponding $H(T)$ in Fig. 18 also contain important information. Only for cyclo-octanol and 2E1H they strongly deviate from unity. For cyclo-octanol, this was attributed to the fact that in plastic crystals the intermolecular coupling mechanisms should differ from those in supercooled liquids [21] but the later findings in 60SN-40GN revealing a factor of 1.05 indicate that this does not seem to be a universal property of plastic crystals [31]. In Ref. [18], the rather small factor for 2E1H was explained considering the clusterlike nature of the relaxing entities, which is believed to cause the Debye relaxation in monohydroxy alcohols [14,125,126].

4.3.2 Fifth harmonic dielectric response

In principle, the susceptibility characterizing the fifth harmonic component of the dielectric response should contain similar information as the third harmonic, discussed in the preceding section. However, as recently pointed out [23], within the theoretical framework of Bouchaud and Biroli [8] $\chi_5$ should be more strongly dependent on the correlation length scale $l$ than $\chi_3$. A critical increase of $l$ is expected within scenarios assuming that the glass transition is due to an underlying phase transition and an approach of amorphous order. Then $\chi_5$ should increase much stronger than $\chi_3$ when approaching the glass transition under cooling [23]. Moreover, the hump should be more pronounced compared to the trivial saturation contribution [23].

As becomes obvious from Fig. 13, the measurement of $\chi_5$ is a non-trivial task due to the small amplitude of the $5\omega$ signals to be detected. However, in a recent joint effort of the group of Ladieu and co-workers and our group [23], it was indeed possible to detect $\chi_5$ for two glass formers, propylene carbonate and glycerol. As an example, Fig. 19 shows spectra of the modulus of $\chi_5$ of propylene carbonate [23]. Similar to $\chi_3$ (Fig. 15), humps are found, preceded by a low-frequency plateau and followed by a continuous decrease at high frequencies. The amplitude of the corresponding humps in the related quantity $X_5^{(5)}$, defined in Eq. (5), was read off after correcting for the trivial contribution from dielectric saturation [23]. In the inset of Fig. 19, it is compared to the corresponding third-order quantity. Just as theoretically predicted for an approach of amorphous order with decreasing temperature, the fifth-order quantity increases much stronger than for the third order. From a detailed comparison of both quantities, in Ref. [23] it was concluded that the cooperative regions are compact and do not have a fractal dimension.

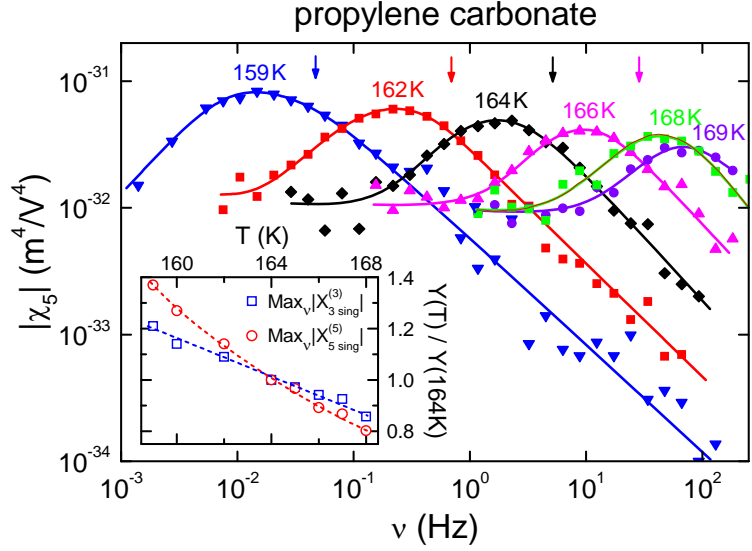

**Fig. 19.** Modulus of the fifth-order harmonic component of the dielectric susceptibility of propylene carbonate measured at various temperatures [23]. The applied field was 380 kV/cm. The arrows indicate the $\alpha$-peak frequencies. The lines are guides to the eyes. The inset shows the amplitude of the hump maximum of the quantities $X_{3\,\text{sing}}^{(3)}$ and $X_{5\,\text{sing}}^{(5)}$ (Eq. (4)), corrected for the trivial saturation contribution and scaled to the value at 164 K. The lines are guides to the eyes.

Finally, in Fig. 20, the linear, third- and fifth-order susceptibilities are compared to each other for glycerol and propylene carbonate. The shown curves are scaled to the value of their low-frequency plateaus, dominated by the trivial response. Obviously, just as predicted within the amorphous-order approach, the non-trivial hump is most pronounced for the fifth-order susceptibility.

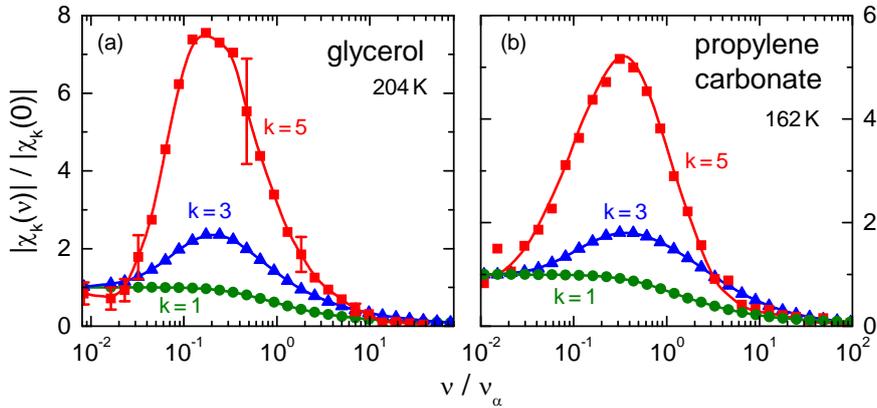

**Fig. 20.** Comparison of fifth-order ($k = 5$), third-order ($k = 3$) and linear ($k = 1$) susceptibilities for glycerol (a) and propylene carbonate (b) at 204 and 163 K, respectively [23]. The applied fields were 780 kV/cm (a) and 380 kV/cm (b). The lines are guides to the eyes.

## 5 Summary and conclusions

The present overview demonstrates the rich variety of phenomena revealed by nonlinear dielectric spectroscopy of glass-forming materials. Some universalities of nonlinear dielectric response can be stated, e.g., the field-induced increase of the dielectric loss at the high-frequency flank of the $\alpha$ relaxation. However, depending on the system class or even for single systems, different dielectric behaviour is also found, e.g., concerning the nonlinearity of high-frequency processes as the excess wing and $\beta$ relaxation. Of special interest are the susceptibilities characterizing the higher-harmonic dielectric response. For all investigated materials, except for the monohydroxy alcohol 1-propanol, they show a spectral shape as predicted by the model by Biroli and Bouchaud assuming a growth of molecular cooperativity and the approach of amorphous order under cooling [8]. However, different explanations were also proposed and further detailed comparisons with the different model predictions are needed to settle controversies concerning the interpretation of such spectra.

The increasing availability of nonlinear dielectric data in recent literature has also triggered the consideration of numerous different, partly contradicting mechanisms for their explanation. It seems somewhat unsatisfactory that often different aspects of nonlinear dielectric response in the same material are interpreted using different models. Clearly more experimental and theoretical efforts are needed trying to at least partly unify these seemingly different approaches.

This work was supported by the Deutsche Forschungsgemeinschaft via Research Unit FOR 1394. Simulating discussions with S. Albert, G. Biroli, U. Buchenau, G. Diezemann, G.P. Johari, F. Ladieu, K.L. Ngai, R.M. Pick, and R. Richert are gratefully acknowledged.